\documentclass{aa}  

\usepackage{graphicx}
\usepackage{txfonts}

\newcommand{\App}[1]{Appendix~\ref{sec:#1}}

\newcommand{\Fg}[1]{Figure~\ref{fig:#1}}

\newcommand{\Tb}[1]{Table~\ref{tab:#1}}

\newcommand{\Welbanks}{\hyperlink{W24}{W24}}
\newcommand{\Sing}{\hyperlink{S24}{S24}}
\newcommand{\Krishnamurthy}{\hyperlink{K25}{K25}}

\usepackage{CJKutf8}
\usepackage[version=4]{mhchem}
\usepackage[colorlinks=true, linkcolor=blue, citecolor=blue, urlcolor=blue]{hyperref}

\usepackage{xcolor,ulem,bm}

\makeatletter
\def\uwave{\bgroup \markoverwith{\lower3.5\p@\hbox{\sixly \textcolor{red}{\char58}}}\ULon}
\font\sixly=lasy6
\makeatother

\begin{document}

   \title{A Cloudy Fit to the Atmosphere of WASP-107 b}

   \author{Helong Huang (\begin{CJK*}{UTF8}{gbsn}黄赫龙\end{CJK*})
          \inst{1}\fnmsep\thanks{Corresponding author: \href{huanghl22@mails.tsinghua.edu.cn}{huanghl22@mails.tsinghua.edu.cn}}
          \and
          Michiel Min\inst{2}
          \and
          Chris W. Ormel\inst{1}
          \and
          Achr\`ene Dyrek\inst{3}
          \and
          Nicolas Crouzet\inst{4,5}
          }

   \institute{Department of Astronomy, Tsinghua University,
              Haidian DS 100084 Beijing, China
         \and
             SRON Netherlands Institute for Space Research, Niels Bohrweg 4, 2333CA Leiden, The Netherlands
        \and
            Space Telescope Science Institute, Baltimore, Maryland, USA
        \and
            Kapteyn Astronomical Institute, University of Groningen, P.O. Box 800, 9700 AV Groningen, The Netherlands
        \and
            Leiden Observatory, Leiden University, P.O. Box 9513, 2300 RA Leiden, The Netherlands
             }

   \date{Received September 15, 1996; accepted March 16, 1997}

  \abstract
  {WASP-107~b has been observed comprehensively by JWST in the near- and mid-IR bands, making it an ideal planet to probe the composition and internal dynamics. Recent analysis reveals a $8-10\ \mathrm{\mu m}$ silicate feature, but it still remains uncertain how silicate clouds form on this planet.}
   {We aim at fitting the complete JWST spectrum of WASP-107~b, from $0.9\ \mathrm{\mu m}$ to $12\ \mathrm{\mu m}$ with a physically motivated cloud model and self-consistent temperature profile.}
   {Two-stream radiative transfer is coupled to a cloud formation model until convergence between cloud and temperature profiles is reached. We search a model grid spanning metallicity, turbulent diffusivity, internal heat flux and nucleation parameters to find the best fit model.}
   {The silicate cloud feature at $10\ \mathrm{\mu m}$ and the near-IR  molecular band strength can be simultaneously and naturally explained without assuming a parametrized temperature profile. A moderate vertical diffusivity of $K_{zz}$=$10^{9}\ \mathrm{cm}^2\mathrm{s}^{-1}$ is needed to bring the cloud particles to the upper atmosphere of WASP-107~b. This $K_{zz}$ is favored by the joint fitting of the near-IR water feature and mid-IR silicate feature -- both sensitive to clouds. From the strength of \ce{H2O} and \ce{CO2} bands, our model suggests a metallicity 17 times solar.}
   {Even in warm planets such as WASP-107~b, silicate clouds can form in the relatively cool upper atmosphere because turbulence uplifts vapor and cloud particles.
    Despite having considerably fewer degrees of freedom, the self-consistent modeling approach successfully fits WASP-107b's multi-wavelength data, instilling confidence in the derived physical parameters.
    }


   \maketitle

\section{Introduction}

WASP-107~b is a warm, super-Neptune-mass planet around a K6 star. 
It has a size of $0.95\ R_\mathrm{Jup}$ and is on a 5.7-day orbit \citep{AndersonEtal2017}.
Follow-up RV measurement refined its mass to $30.5\pm 1.7\ M_\oplus$ \citep{PiauletEtal2021}.
WASP-107~b's large radius makes it an ideal target to observe the transmission spectrum, because the large planet-to-star radius ratio magnifies the spectral features. 
The HST transmission spectrum of this planet shows water feature and cloud extinction in the near-IR \citep{KreidbergEtal2018}.
Recently, JWST NIRCam and NIRSpec has found \ce{H2O}, \ce{CO2}, \ce{SO2}, \ce{CO}, \ce{NH3} and \ce{CH4} on WASP-107~b with high confidence levels \citep[][\hypertarget{W24}{W24} and \hypertarget{S24}{S24} hereafter]{WelbanksEtal2024, SingEtal2024}.
Furthermore, both JWST and HST have found \ce{He} absorption around the transit, indicating ongoing atmospheric escaping process on this planet \citep[][\hypertarget{K25}{K25} hereafter]{SpakeEtal2018, KrishnamurthyEtal2025}.

With the observations from JWST, clouds on WASP-107~b are found at all wavelengths from $2.4\ \mathrm{\mu m}$ to $12\ \mathrm{\mu m}$.
\Sing\  showed that clouds are needed to lift the transit depth between $3.6$ and $4.0\ \mathrm{\mu m}$.
\citet{DyrekEtal2024} found that the JWST/MIRI observation clearly favors the presence of silicate clouds, hinted by the $10\ \mathrm{\mu m}$ \ce{Si-O} stretching feature. 
Their retrieval suggested the presence of submicron sized cloud particles at the ${\sim}10^{-4}\ \mathrm{bar}$ level.
However, at those cool layers, it is believed that any silicates would have rained out to deeper, hotter region.
On the other hand, from the depletion of \ce{CH4}, it has been found that the atmosphere of WASP-107~b is characterized by strong turbulence (\Sing), which could potentially transport particles up into the visible atmosphere.

Most previous works fit the transmission spectrum of WASP-107~b with parameterized cloud properties. For example, \Sing\  applied gray clouds and a power-law haze opacity to retrieve the $2.6$ to $5.2\ \mathrm{\mu m}$ NIRSpec spectrum. \citet{DyrekEtal2024} used realistic optical properties of silicates, but parameterized the vertical distribution of clouds. Also, \Welbanks\ used an additional opacity with skewed Gaussian shape along the wavelength dimension to mimic the mid-IR silicate opacity. Even though these cloud parameterizations provided successful fits to the spectra, the question how silicate clouds form in the atmosphere of WASP-107~b has not yet been addressed.

Therefore we aim to investigate whether the cloud features on WASP-107~b can be understood with a realistic cloud formation model.
Instead of parameterizing cloud properties, we simulate the cloud formation on WASP-107~b with the physically-motivated cloud model \texttt{ExoLyn} \citep{HuangEtal2024}.
This allows us to identify which silicate species the cloud particles are composed of.
In addition, the temperature is calculated with a two-stream radiative transfer model that accounts for both molecular and cloud opacities (Huang et al., in prep.). 
Our model results are compared with the transmission spectrum taken by  JWST/NIRISS (\Krishnamurthy), NIRCam (\Welbanks) and MIRI \citep{DyrekEtal2024}, covering visible to mid-IR wavelengths.
Data at wavelengths $\lambda<1\ \mathrm{\mu m}$ and $2{-}2.5\ \mathrm{\mu m}$ are excluded, as they may be affected by the transit light source effect (\Krishnamurthy).
The NIRCam data, from $2.5\ \mathrm{\mu m}$ to $5\ \mathrm{\mu m}$, used in this study does not show evidence for starspot crossing event (\Welbanks).
Beyond $\lambda>2.5\ \mathrm{\mu m}$ the effect of starspots is decreasing with wavelength \citep{RackhamEtal2018}. 
Therefore, the effect of transit light source on our findings is minor.

\section{Results} \label{sec:results}
\begin{table}
    \caption{Parameters of the grid search. The highlighted ones indicate the best-fit.}
    \small
    \begin{tabular}{l|lllll|l}
        \hline
        Parameter       & \multicolumn{5}{c}{grid values} & Unit \\
        \hline
        $\log Z/Z_\odot$ & 0.5 & 0.75 & 1 & \textbf{1.25} & 1.5 & \\
        $K_{zz}$ & $10^{8}$ & $10^{8.5}$ & $\bm{10^{9}}$ & $10^{9.5}$ & $10^{10}$  & $\mathrm{cm}^2\mathrm{s}^{-1}$ \\
        $T_\mathrm{int}$ & 350 & 450 & \textbf{550} & - & - & $\mathrm{K}$\\
        $\dot{\Sigma}_\mathrm{nuc}$    & $10^{-21}$ & \bm{$10^{-20}$} & $10^{-19}$ & $10^{-18}$ & - & $\mathrm{g\,cm}^{-2}\mathrm{s}^{-1}$ \\
        $P_\mathrm{nuc}$ & $10^{-4}$ & $\bm{10^{-3}}$ & $10^{-2}$ & $10^{-1}$ & - & $\mathrm{bar}$\\
        \hline
    \end{tabular}\label{tab:parameters}
\end{table}

We couple the cloud model \texttt{ExoLyn} with two-stream radiative transfer to get a self-consistent TP profile and cloud structure. Our model is summarized in the Appendix \ref{sec:appendix} and the details are described in \citet{HuangEtal2024} and a companion paper. 
We construct a grid of parameters (\Tb{parameters}), with varying metallicity $Z$, vertical turbulent diffusivity $K_{zz}$, effective temperature of the internal emission flux $T_\mathrm{int}$, nuclei injection rate $\dot{\Sigma}_\mathrm{nuc}$ and nucleation depth $P_\mathrm{nuc}$ to find the parameters that best match the observations, including the mid-IR silicate features and near-IR molecular band strength.
For each parameter combination, we adjust the equilibrium gas composition to account for the vertical mixing and sulfur photochemistry, which are described in the Appendix \ref{sec:diseq-adjust}.
The observational data combining JWST/NIRISS (\Krishnamurthy), JWST/NIRCam and JWST/MIRI presented in \Welbanks, covers a broad wavelength range between $0.85$--$12\ \mathrm{\mu m}$.
The best fit model is found with $Z=17\ Z_\odot$, $K_{zz} = 10^{9}\ \mathrm{cm}^2\mathrm{s}^{-1}$, $T_\mathrm{int}=550\ \mathrm{K}$, $P_\mathrm{nuc} = 10^{-3}\ \mathrm{bar}$ and $\dot{\Sigma}_\mathrm{nuc}=10^{-20}\ \mathrm{g}\ \mathrm{cm}^{-2}\ \mathrm{s}^{-1}$. 

\begin{figure}
    \centering
    \includegraphics[width=\columnwidth]{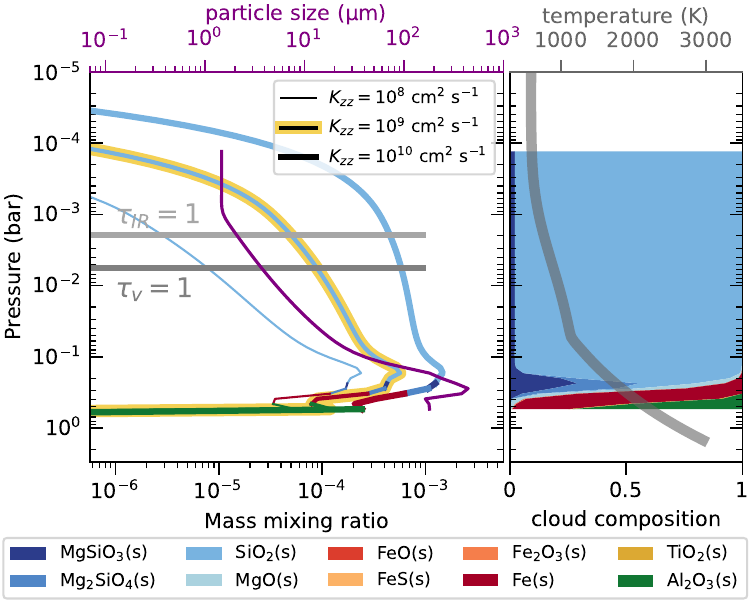}
    \caption{Self-consistent temperature and cloud profiles, resulting from the joint cloud and radiative transfer model. Left: cloud profiles obtained for three $K_{zz}$ values. The best fit model has $K_{zz}=10^{9}\ \mathrm{cm}^2\mathrm{s}^{-1}$ (the line with golden highlight) and the corresponding particle size is shown by the purple line. The horizontal gray lines indicates the visible and IR photosphere. Right: cloud composition by mass fraction of the best-fit model. The gray line shows the atmosphere temperature profile of the best fit.}
    \label{fig:clouds}
\end{figure}

\begin{figure*}
    \centering
    \includegraphics[width=\textwidth]{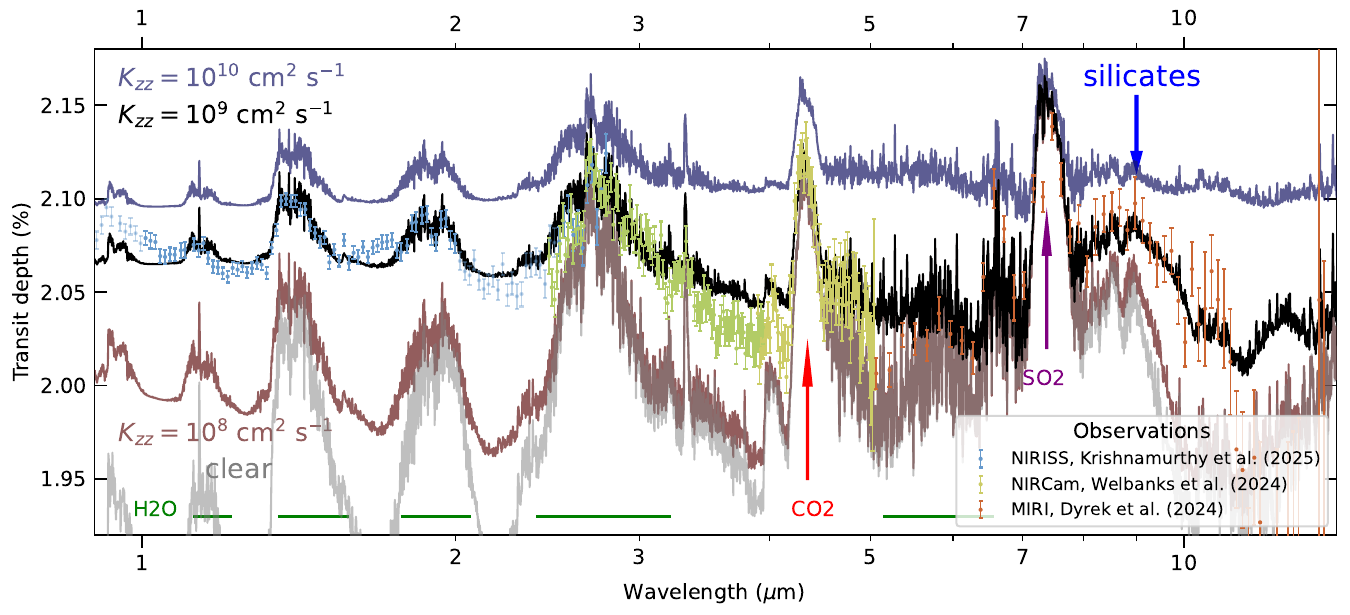}
    \caption{JWST observations of WASP-107~b (colored vertical bars) along with the best fit model (black line; $K_\mathrm{zz}=10^{9}\ \mathrm{cm}^2\mathrm{s}^{-1}$), resulting in a $\tilde{\chi}^2_\mathrm{red}=3.4$. The NIRISS observations are shifted vertically by $130$ ppm to match the NIRCam data at the wavelengths where they overlap. The runs with higher and lower $K_{zz}$ are shown with blue and brown lines, respectively, to illustrate the spectroscopic effects of varying the turbulent diffusivity. The simulated spectrum of a cloudless atmosphere is plotted in gray.}
    \label{fig:spectrum}
\end{figure*}

The converged TP profile and the cloud structure are shown in \Fg{clouds}.
For the best-fit $K_{zz} = 10^{9}\ \mathrm{cm}^2\mathrm{s}^{-1}$ run, the upper cloud layer is dominated by \ce{SiO2}, until $0.3$ bar, where the dominant species changes to \ce{MgSiO3}, \ce{Mg2SiO4} and \ce{Fe} because the temperature is high enough for \ce{SiO2}  to evaporate. 
Due to gravitational settling, the cloud mixing ratio peaks at $0.2\ \mathrm{bar}$. Clouds evaporate at a pressure of $0.6\ \mathrm{bar}$ where $T\approx2200\ \mathrm{K}$. At $2\times 10^{-3}\ \mathrm{bar}$ and $5\times 10^{-3}\ \mathrm{bar}$ the atmosphere becomes optically thick to the stellar IR and visible photons. 
Here, the cloud particles reach sizes of ${\sim}\mu\mathrm{m}$.
Increasing $K_{zz}$ does not change the dominant cloud species (\ce{SiO2}), but does lead to more vertically extended clouds.
This is because vapor gets replenished more efficiently in the cloud forming region and the cloud particles undergo stronger updraft motion.
The photospheric temperature we obtain ($730\ \mathrm{K}$) agrees with the retrieved values in \Krishnamurthy\  (${\sim}700\ \mathrm{K}$) and lies in between the radiative-convective-equilibrium models presented by \Welbanks\  (${\sim}500\ \mathrm{K}$) and \Sing\  (${\sim}800\ \mathrm{K}$).

\Fg{spectrum} shows the synthetic transmission spectrum compared to observations.
We also plot the simulation results obtained by varying the vertical turbulent diffusivity $K_{zz}$ with respect to the best-fit model to demonstrate how this parameter affects the silicate features in the spectrum.
The black solid line shows the transmission spectrum corresponding to the best-fit $K_{zz}=10^{9}\ \mathrm{cm}^2\mathrm{s}^{-1}$ run. 
Our model provides a good match to the strength of both the near-IR molecular bands and the mid-IR silicate feature.
Compared to the clear spectrum, \ce{SiO2} clouds give rise to enhanced absorption at $8$ to $10\ \mathrm{\mu m}$.
In the near-IR, scattering by $\mu$m-sized particles raises the transit depth at wavelengths below $2.5\ \mathrm{\mu m}$, suppressing the $1.2$, $1.5$, $1.9$ and $2.8\ \mathrm{\mu m}$ \ce{H2O} features. Besides \ce{H2O}, other molecular features are also reproduced. The atmospheric metallicity affects the strength of the $4.4\ \mathrm{\mu m}$ \ce{CO2} feature, and is constrained at ${\approx}17\ Z_\odot$. 
Our model also reproduces the \ce{SO2} feature at $7.4\ \mathrm{\mu m}$ under the assumption of a $9\times 10^{-3}$ conversion ratio from \ce{H2S} to \ce{SO2}.
This conversion is applied to mimic the effects of photochemistry (not included in our model), which is believed to be the cause of the presence of \ce{SO2} on hot Jupiters \citep[][\Welbanks]{TsaiEtal2023}. 

Varying the turbulent diffusivity parameter leads to vertical shifts in the transmission spectrum. 
When $K_{zz} = 10^{10}\ \mathrm{cm}^2\mathrm{s}^{-1}$, clouds are more extended. In this case, cloud extinction ${<}2\ \mathrm{\mu m}$ is too strong and the transit depth is overestimated compared to the  observations.
On the other hand, when $K_{zz} = 10^8\ \mathrm{cm}^2\mathrm{s}^{-1}$, clouds settle too deep, leading to a nearly invisible silicate feature and larger near-IR water line amplitudes.
In conclusion, the cloud features at $1{-}4\ \mathrm{\mu m}$ and $8{-}10\ \mathrm{\mu m}$ can be reproduced only with a turbulent diffusivity $K_{zz}=10^{9}\ \mathrm{cm}^2\mathrm{s}^{-1}$, consistent with the value inferred from the gas-phase disequilibrium chemistry model used in \Welbanks\  and \citet{ChangeatEtal2025}.

\begin{figure*}
    \centering
    \includegraphics[width=\textwidth]{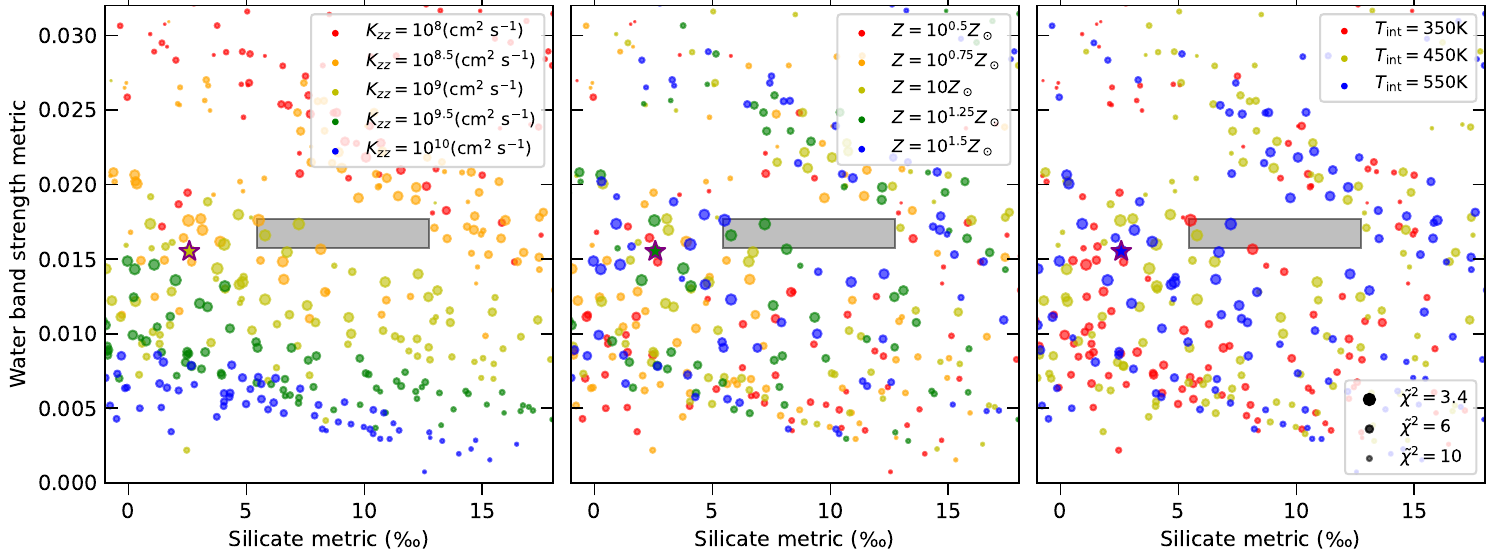}
    \caption{Scatter plots demonstrating how the near-IR \ce{H2O} line strength and the $10\ \mathrm{\mu m}$ silicate metric depend on the diffusivity parameter ($K_{zz}$; left), metallicity (Z; middle) and internal energy flux ($T_\mathrm{int}$; right) model parameters. The size of the dots corresponds to the $\tilde{\chi}^2$ of their fits to the JWST observations. The gray rectangle corresponds to the JWST observational data and their $1\sigma$ uncertainties. The best-fit model run (star), is in $2\sigma$ agreement with the observation.}
    \label{fig:scatter}
\end{figure*}

In \Fg{scatter} we show how the $K_{zz}$, $T_\mathrm{int}$ and $Z$ parameters change the spectroscopic features that are affected by clouds. 
To analyze the clouds' ability to mute molecular lines, we measure the strength of the $1.4\ \mathrm{\mu m}$ \ce{H2O} band.
We take the water line strength metric in \citet{Stevenson2016}:
\begin{equation}
    f_\mathrm{water} = (\eta^\mathrm{water} - \eta^\mathrm{cont})/\eta^\mathrm{cont}, \label{eq:metric}
\end{equation}
where $\eta^\mathrm{water}$ is the transit depth at the peak of the \ce{H2O} band (i.e., $1.36-1.44\ \mathrm{\mu m}$) and $\eta^\mathrm{cont}$ is that at the continuum wavelengths (i.e., $1.22-1.3\ \mathrm{\mu m}$).
In addition to $f_\mathrm{water}$, we quantify the strength of the $10\ \mathrm{\mu m}$ silicate feature through a silicate metric $f_\mathrm{silicate}$, obtained by decomposing the transmission spectrum from $8.4$--$10.1\ \mathrm{\mu m}$ into Legendre polynomials up to second order: $\eta(\lambda) = L_0+L_1P_1(\lambda)+L_2P(\lambda)$. Then
\begin{equation}\label{eq:f-silicate}
    f_\mathrm{silicate} = (L_1^\ast-2L_2^\ast)/L_0.
\end{equation}
where "$^\ast$" denote that $L_1$ is limited to the range $[-0.01L_0,0]$ and $L_2<0$, respectively.
The two terms encapsulate how the silicate feature manifests in the spectrum. Without silicates, the molecular opacity steeply decreases from $8$--$10\ \mathrm{\mu m}$ and $f_\mathrm{silicate}$ (through $L_1$) evaluates negatively. 
Silicate absorption reduces this decline: it flattens the opacity slope and increases $f_\mathrm{silicate}$. 
With further increase of cloud thickness, a distinct absorption bump near $9\ \mathrm{\mu m}$ emerges, captured by $L_2$.
Defined in this way, the silicate metric indicates substantial evidence for silicates when $f_\mathrm{silicate}$ evaluates ${\gtrsim}0.01$. 
For WASP-107 b, the slope contribution ($L_1$) dominates and $f_\mathrm{silicate}=0.009$, indicating the likely presence of silicates.

Although a large scatter is present in \Fg{scatter} due to the variation of parameters, a decreasing trend of $f_\mathrm{water}$ and slightly increasing trend of $f_\mathrm{silicate}$ is identified with increasing $K_{zz}$. 
The stronger turbulence dredges up more vapor to the upper atmosphere, where it condenses to form silicates and to suppress the molecular features.
Increasing the metallicity directly leads to a higher \ce{H2O} abundance and stronger \ce{H2O} features.
On the other hand, the metallicity and $T_\mathrm{int}$ parameter does not influence $f_\mathrm{water}$ and $f_\mathrm{silicate}$ in a systematic way. However, a hotter interior promotes quenching of \ce{CH4} in the upper atmosphere, resulting in better fits to JWST observations (as indicated by the size of the points) : the average $\tilde{\chi^2}$ of the top 10 best-fit runs decreases from $4.3$ to $3.8$ when increasing $T_\mathrm{int}$ from $350\ \mathrm{K}$ to $550\ \mathrm{K}$.
There are 3 models lying within $1\sigma$ from the observational constraints on the silicate metric and the water metric. These points, all with $\tilde{\chi^2}<3.7$, are statistically indistinguishable with the best-fit simulation. They all have parameters $K_{zz}\approx 10^9\ \mathrm{cm}^2\mathrm{s}^{-1}$ and $Z\approx 17$, confirming the robustness of our inferences.

Finally, the cloud profiles (and associated metrics) are mostly insensitive to nucleation parameters (not plotted). Higher nucleation rates are neutralized by coagulation while turbulence efficiently spreads cloud nuclei vertically \citep{HuangEtal2024}. 

\section{Conclusions and Discussion}
The cloud characteristics of WASP-107~b have been investigated using a model strategy that couples a two-stream radiative transfer model and a physically-motivated cloud formation model in a self-consistent manner. Our main findings are as follows:
\begin{enumerate}
    \item The near-IR  molecular band strength observed by  JWST/NIRISS and NIRCam and the $10\ \mathrm{\mu m}$ silicate feature observed by JWST/MIRI can be reproduced simultaneously without assuming a parametrized cloud profile or temperature profile. Although the temperature in the upper atmosphere  is relatively low compared to the condensation temperature of silicates, vertical updrafts  lifts cloud particles into the observable upper region.
    \item The best-fit model employs a vertical turbulent diffusivity $K_{zz}=10^{9}\ \mathrm{cm}^2\mathrm{s}^{-1}$, a value typical in exoplanet atmospheres \citep{KawashimaMin2021, BaxterEtal2021, BaratEtal2025}. When $K_{zz}$ is lower, clouds become invisible due to gravitational settling, while higher $K_{zz}$ dredges up too much cloud material, inconsistent with the relatively weak silicate feature.
    \item Turbulence governs both the amplitude of the $10\ \mathrm{\mu m}$ silicate feature and the strength of the near-IR water lines. 
        Self-consistently modeling the effect of turbulence on clouds (thickness, intensity) allows us to put tight constraints on $K_{zz}$.
\end{enumerate}

The model used in this work eliminates excessive degrees of freedom compared to models adopting parametrized cloud opacity or TP profile. Despite consisting of just 1200 sets of parameters (\Tb{parameters}), it is able to match the most prominent cloud features. With more and more JWST observations suggesting the presence of clouds on exoplanets \citep{GrantEtal2023, InglisEtal2024}, the method could be applied in future (grid) retrievals. 

Nevertheless, the 1D nature of the model is an assumption. Recently, differences in the morning vs evening limb spectrum haven been pointed out on WASP-107~b \citep{MurphyEtal2024}. The evening limb shows a stronger $2.7\ \mathrm{\mu m}$ \ce{H2O} feature and a higher transit depth, which suggests an atmosphere $100\ \mathrm{K}$ hotter than the morning limb.
The relatively weak $10\ \mathrm{\mu m}$ silicate feature therefore may only originate from the evening limb, as silicate clouds would likely  rain out in the cooler morning limb. The limb asymmetry likely also affects the photochemistry related to the \ce{SO2} feature, as \ce{SO2} can only form on the day side.
A future effort would therefore be to integrate two limb models into our code, with different temperatures and compositions.

With its large radius for its mass, the super-puff WASP-107 b occupies a distinctive place in the exoplanet population. Similar to previous studies (\Sing\ and \Welbanks), our modeling of the planet's atmospheric indicates a high internal temperature $T_\mathrm{int}$, suggesting active internal heating, driven by, e.g, tidal or Ohmic dissipation \citep{Batygin2025, TremblinEtal2017}. Notably, our inferred atmospheric metallicity of 17x solar is consistent with the constraint in \Welbanks\ but lower than the value reported by \Sing\ -- a finding that results from our self-consistent cloud model (see \App{highZ}). The slightly lower atmospheric metallicity leaves ample room for the bulk of the heavy elements to reside deep within the planet's interior.

\begin{acknowledgements}
      This work was supported by the National Natural Science Foundation of China (NSFC) project
      no. 12473065 and 12233004.
      The authors thank Kazumasa Ohno and Sharon Xuesong Wang for insightful discussion, and the anonymous referee whose comments substantially improved the quality of the paper.
\end{acknowledgements}

\bibliographystyle{aa}
\bibliography{ads}

\begin{appendix}
\section{Methods}\label{sec:appendix}

In this work we compute the self-consistent cloud structure and temperature profile by coupling the multi-species cloud code \texttt{ExoLyn} with a two-stream radiative transfer method. 
Here we briefly summarize the physical principles of our simulations. Details of our radiative transfer model are described in the companion paper.

\subsection{Coupling between cloud formation and radiative transfer}

\texttt{ExoLyn} \citep{HuangEtal2024} is a 1D cloud formation model that solves for the cloud composition, number density and size of the cloud particles.
The cloud formation in \texttt{ExoLyn} is simulated as a chemical kinetics process instead of chemical equilibrium. 
For the cloud forming material, we assume \ce{C} combines into \ce{CO} and the leftover \ce{O} forms \ce{H2O}, which applies for the hot deep atmosphere. Due to turbulence (eddy diffusion), molecules are transported to cooler, upper layers and condense into clouds.
The reaction rates at which condensates form are evaluated using the local vapor concentration and the thermo-chemical property of the cloud forming reactions. 
Once formed, a cloud particle is subject to sedimentation, turbulent diffusion and coagulation, until it evaporates at the bottom of the atmosphere.
The steady state cloud profile is solved using a computationally efficient relaxation method.
During the cloud formation process, gas-phase chemical reactions are neglected, for simplicity. After a cloud structure has been obtained, the remaining gas species are forced to chemical equilibrium using \texttt{FastChem} \citep{StockEtal2018, StockEtal2022}.

\begin{table}
    \caption{opacity data used in this work}
    \label{tab:opa_data}
    \centering
    \small
    \begin{tabular}{l l}
    \hline
    Opacity source & Reference \\
    \hline
    \ce{Al2O3}(s)  &   \citet{KoikeEtal1995} \\
    \ce{Fe}(s), \ce{MgO}(s)     &   \citet{Palik1991} \\
    \ce{Fe2O3}(s) & A.H.M.J. Triaud\tablefootmark{a} \\
    \ce{Mg2SiO4}(s), \ce{MgSiO3}(s)  & \citet{JagerEtal2003} \\
    \ce{FeO}(s) & \citet{HenningEtal1995} \\
    \ce{FeS}(s) & \citet{PollackEtal1994} \\ 
    \ce{H2O}, \ce{CO} & \citet{RothmanEtal2010} \\
    \ce{CO2} & \citet{YurchenkoEtal2020} \\
    \ce{CH4} & \citet{YurchenkoEtal2017}\\
    \ce{H2S} & \citet{AzzamEtal2016}\\
    \ce{Na} & \citet{AllardEtal2019}\\
    \ce{K} & \citet{MolliereEtal2019}\\
    \ce{Mg}, \ce{Fe}, \ce{Al} & K. Molaverdikhani\tablefootmark{b}\\
    \ce{SiO} & \citet{YurchenkoEtal2022}\\
    \ce{TiO} & \citet{McKemmishEtal2019}\\
    \ce{NH3} & \citet{ColesEtal2019}\\
    \ce{H2}-\ce{H2} & \citet{BorysowEtal2001, Borysow2002}\\
    \ce{H2}-\ce{He} & \citet{BorysowEtal1988, BorysowEtal1989};\\
    &\citet{BorysowFrommhold1989}\\
    \hline
    \end{tabular}
    \tablefoot{
    \tablefoottext{a}{\url{https://www.astro.uni-jena.de/Laboratory/OCDB/mgfeoxides.html}}
    \tablefoottext{b}{\url{http://kurucz.harvard.edu/}}
    }
\end{table}

\begin{figure*}
    \centering
    \includegraphics[width=\textwidth]{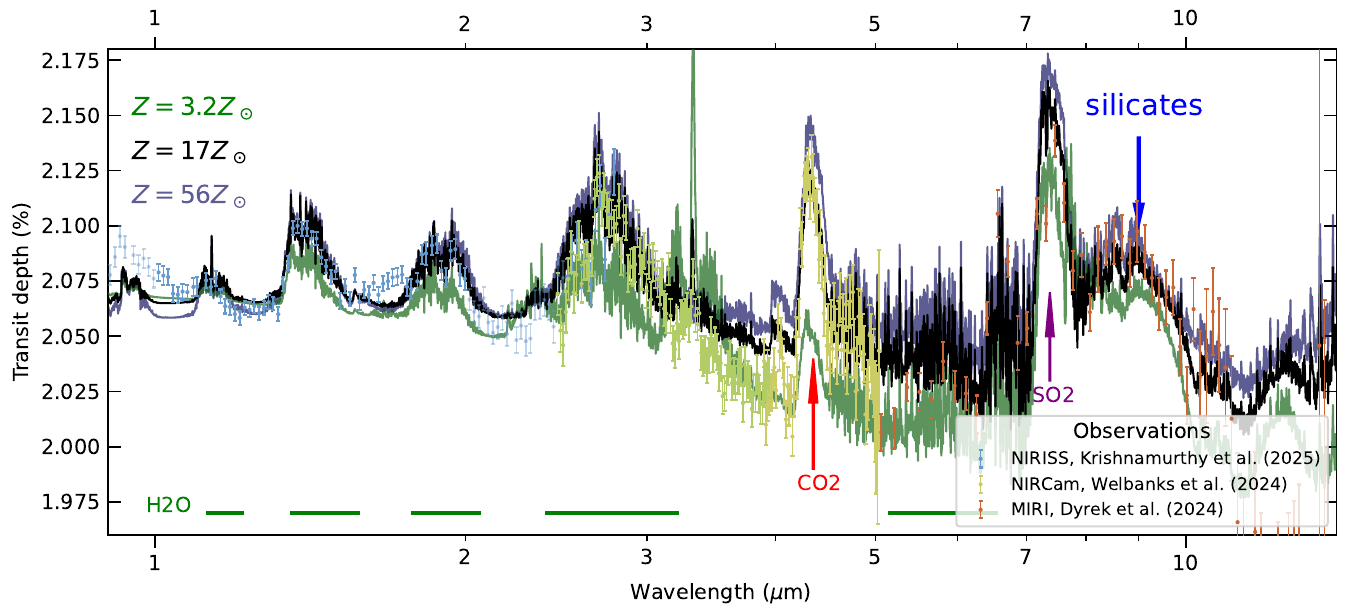}
    \caption{Same as \Fg{spectrum}, but with metallicity $Z=56\ Z_\odot$ and $Z=3.2\ Z_\odot$. For each metallicity, the control parameters (\App{diseq-adjust}) are kept the same as the best-fit model.}
    \label{fig:spectrum_highZ}
\end{figure*}

Cloud formation depends sensitively on the temperature structure of the atmosphere. 
Therefore instead of leaving the temperature structure as free parameters, we seek to solve it self-consistently with cloud formation.
The first step in radiative transfer calculation is to compute the cloud and gas phase opacity structure. For the cloud opacity, we take the same setting as \citet{HuangEtal2024}. 
Effective medium theory (Bruggeman rule) is applied to the composition of the cloud particles to get the average (effective) refractive index of the material mixture. 
Then \texttt{Optool} \citep{DominikEtal2021} is run to generate the cloud absorption and scattering opacity, assuming distribution of hollow spheres \citep[DHS, ][]{MinEtal2005} as the shape for the cloud particles .
We use \texttt{Exo-k} \citep{LeconteEtal2021}  to mix the correlated-k tables of the individual gas species on the wavelength range from $0.3\ \mathrm{\mu m}$ to $50\ \mathrm{\mu m}$, with a spectral resolution $R=300$. 
The gas opacity sources included in the simulations are \ce{H2O}, \ce{CH4}, \ce{CO}, \ce{CO2}, \ce{NH3}, \ce{H2S}, \ce{Na}, \ce{K}, \ce{Mg}, \ce{Fe}, \ce{Al}, \ce{SiO}, \ce{TiO} and \ce{H2}-\ce{H2}, \ce{H2}-\ce{He} collisional induced absorption, summarized in \Tb{opa_data}.

Once the opacity is computed, we generalize the two-stream radiative transfer method in \citet{Guillot2010} to get the temperature profile.
The radiation field is decomposed into two components \citep{ToonEtal1979, MeadorWeaver1980} -- downward irradiation from the star in visible band and the upward irradiation from the planet interior in IR band.
For both the IR and visual components, moment equations \citep{Mihalas1978} are solved, with the addition of Eddington approximation as closure condition \citep{Eddington1916}.
The location where the stellar energy is deposited is determined by the opacity distribution in the atmosphere, especially clouds, because they tend to absorb stellar visual photons efficiently. 
Applying energy balance -- the radiation energy received by a layer equals to the blackbody radiation it emits -- the temperature at a certain layer is computed.
The assumptions and simplifications in the above method is benchmarked against the radiative transfer code \texttt{PICASO} in the companion paper. 
The temperature gradient computed from radiation transfer becomes super-adiabatic in the deep atmosphere ($\gtrsim 0.1\ \mathrm{bar}$). In reality, convection transports the heat more efficiently in these region and homogenizes the temperature. We correct the super-adiabatic temperature gradient following the method presented in \citet{MalikEtal2019}, which guarantees that the temperature gradient is not larger than $\nabla_\mathrm{ad}$ (assumed to be $2/7$) and the net energy flux is conserved throughout the atmosphere.

In this work, \texttt{ExoLyn} and the radiative transfer code are coupled iteratively. Starting from a guess temperature profile, we compute the clouds structure and then the gas and cloud opacity.
After this, the temperature corresponding to the opacity structure is computed to update the guess temperature.
This process is run iteratively until convergence in temperature is reached.
In the end, we generate the transmission spectra with \texttt{petitRADTRANS} \citep{MolliereEtal2019}.

\subsection{Adjustment to the equilibrium chemistry} \label{sec:diseq-adjust}

Before computing the transmission spectrum, we made several adjustments to the gas profile under chemical equilibrium.
It is found that the gas composition on WASP-107~b is affected by disequilibrium chemistry such as photochemistry and quenching (\Welbanks). 
For example, the JWST transmission spectrum of WASP-107~b shows the signal of \ce{SO2}, in contrast to \ce{H2S} predicted by chemical equilibrium.
The \ce{SO2} could be generated from oxidation of \ce{H2S} in photochemical reactions.
However, as the main focus of this work is to understand the cloud features on WASP-107~b with a microphysical model, running a photochemistry model is beyond the scope of the model.
Therefore, we artificially convert some fraction of \ce{H2S} to \ce{SO2}.

In addition, at the temperature akin to WASP-107~b, the presence of \ce{CH4} and the absence of \ce{NH3} is expected by equilibrium chemistry. 
However, the observed underabundance of \ce{CH4} and overabundance of \ce{NH3} are signposts of disequilibrium chemistry, specifically vertical quenching.
Under strongly turbulent condition, \ce{CH4}-poor gas from the deep atmosphere is dredged up into the visible, upper atmosphere, and vice versa, reducing the concentration of \ce{CH4}. As simulating disequilibrium chemistry is beyond the scope of this work, we mimic the above effects by postprocessing the gas profile before generating the transmission spectrum:
\begin{itemize}
    \item We convert a fraction of $f_{\ce{H2S}}$ \ce{H2S} to \ce{SO2}.
    \item We replace all molecular abundances above a quenching pressure $P_\mathrm{quench}$ with those values at that location.
\end{itemize}
For each parameter set in \Tb{parameters}, the best-fit $f_{\ce{H2S}}$, $P_\mathrm{quench}$ and the reference pressure $P_\mathrm{ref}$ at which the planet radius is measured are optimized by the conjugate gradient method.
The best fit parameters in \Fg{spectrum} have $f_{\ce{H2S}}=9 \times 10^{-3}$, $P_\mathrm{quench}=0.06\ \mathrm{bar}$. 

Our model manages to fit the $7.2\ \mathrm{\mu m}$ \ce{SO2} feature, but undershoot the $4\ \mathrm{\mu m}$ \ce{SO2} feature. 
The reason could be that the cloud extinction in our model is slightly stronger than it is suggested by the $4\ \mathrm{\mu m}$ \ce{SO2} feature.
Therefore, tuning a smaller $K_{zz}$ could lead to a slightly clearer atmosphere and decrease the transit depth redward to the \ce{SO2} feature, providing a better fit.
Note that the \ce{SO2} concentration in this work is calculated by simply converting 0.6\% of \ce{H2S} to \ce{SO2}, translating to a nearly constant volume mixing ratio ${\approx}{10^{-6}}$ at all locations in atmosphere.
\Welbanks\  applied a 1D radiative-convective–photochemical equilibrium (RCPE) model, which found that \ce{SO2} only form around a level of $10^{-5}\ \mathrm{bar}$, with the peak volume mixing ratio $10^{-4}$. 
Explaining the presence of \ce{SO2} with a cloudy RCPE model is left to a future work.

\subsection{Higher metallicity simulations} \label{sec:highZ}
The best-fit atmospheric metallicity in our work, $Z=17\ Z_\odot$, is consistent with the constraint from \Welbanks\ ($10{-}18\ Z_\odot$) and with the \ce{K} abundance from \Krishnamurthy\ ($8^{+26}_{-6}\ Z_\odot$). However, it is lower than the values reported by \Sing\ ($43\pm8\ Z_\odot$). Motivated by their constraint, we simulate a metal-rich ($Z=56\ Z_\odot$) and a low metallicity ($Z=3.2\ Z_\odot$) model to understand how metallicity affects the spectrum. The control parameters described in \App{diseq-adjust} are  kept the same as those of the best-fit model for each metallicity. The results are shown in \Fg{spectrum_highZ}.

With a $\tilde{\chi}^2=6.5$, the metal-rich atmosphere with $Z=56\ Z_\odot$ offers a worse fit. Although a higher metallicity of $56\ Z_\odot$ elevates the mean molecular weight to $3.07$, its effect on the scale height is not strong enough to suppress the gas and cloud features. The amplitude of the \ce{H2O} features at $1.4\ \mathrm{\mu m}$ and the \ce{CO2} feature at $4.3\ \mathrm{\mu m}$ are overestimated compared to the observations. Increasing the metallicity results in a higher \ce{H2O} abundance and enhances the corresponding features in the near-IR band. The \ce{CO2} abundance is especially sensitive to metallicity, with its signal at $4.3\ \mathrm{\mu m}$ enhanced from 400 ppm for $3.2\ Z_\odot$ to 900 ppm for $56\ Z_\odot$. The silicate feature at $10\ \mathrm{\mu m}$ is not sensitive to metallicity. Although more vapor condenses in metal-rich atmosphere, the silicate-bearing cloud particles gravitationally settle to optically opaque layers once the particles grow larger.

The unusual large radius of WASP-107~b requires a large envelope mass fraction and small core mass of $5M_\oplus$ \citep{PiauletEtal2021}. However, this poses a question to the formation theory how such low mass core accreted an envelope of $25M_\oplus$. 
The strong internal heating ($T_\mathrm{int}$) inferred in our model, \Welbanks\  and \Sing, suggests a hotter envelope and helps to explain the inflated radius of the planet without involving a small core mass fraction. 
This intense interior radiation potentially stems from the tidal heating related to the planet's moderate eccentricity $e\sim 0.05$ \citep{PiauletEtal2021, MurphyEtal2024, YeeVissapragada2025} or from the Ohmic dissipation in the planet interior \citep{Batygin2025}.

\end{appendix}
\end{document}